\documentclass[journal=jpccck,manuscript=article]{achemso}
\setkeys{acs}{maxauthors=10} 
\setkeys{acs}{articletitle = true} 

\usepackage{chemformula} 
\usepackage[T1]{fontenc} 
\usepackage{amsmath,amssymb}
\usepackage{bm}
\usepackage{color}
\usepackage{ulem} 

\newcommand*\mycommand[1]{\texttt{\emph{#1}}}

\author{Goro Shibata}
\affiliation{Department of Physics, The University of Tokyo, Tokyo, Japan}
\alsoaffiliation{Department of Applied Physics, Tokyo University of Science, Tokyo, Japan}
\email{shibata@rs.tus.ac.jp}
\phone{+81(3)-5876-1717}
\author{Choongjae Won}
\affiliation{Max Planck POSTECH/Korea Research Initiative, Pohang University of Science and Technology, Pohang, Korea}
\alsoaffiliation{Laboratory of Pohang Emergent Materials, Pohang Accelerator Laboratory, Pohang, Korea}
\author{Jaewook Kim}
\affiliation{Department of Physics and Astronomy, Rutgers University, Piscataway, New Jersey, USA}
\author{Yosuke Nonaka}
\affiliation{Department of Physics, The University of Tokyo, Tokyo, Japan}
\author{Keisuke Ikeda}
\affiliation{Department of Physics, The University of Tokyo, Tokyo, Japan}
\author{Yuxuan Wan}
\affiliation{Department of Physics, The University of Tokyo, Tokyo, Japan}
\alsoaffiliation{The Institute for Solid State Physics, The University of Tokyo, Kashiwa, Japan}
\author{Masahiro Suzuki}
\affiliation{Department of Physics, The University of Tokyo, Tokyo, Japan}
\author{Tsuneharu Koide}
\affiliation{Institute of Materials Structure Science, 
High Energy Accelerator Research Organization, Tsukuba, Ibaraki, Japan
}
\author{Arata Tanaka}
\affiliation{Department of Quantum Matter, Hiroshima University, Hiroshima, Japan}
\author{Sang-Wook Cheong}
\affiliation{Department of Physics and Astronomy, Rutgers University, Piscataway, New Jersey, USA}
\alsoaffiliation{Max Planck POSTECH/Korea Research Initiative, Pohang University of Science and Technology, Pohang, Korea}
\alsoaffiliation{Laboratory of Pohang Emergent Materials, Pohang Accelerator Laboratory, Pohang, Korea}
\author{Atsushi Fujimori}
\affiliation{Department of Physics, The University of Tokyo, Tokyo, Japan}
\alsoaffiliation{Department of Applied Physics, Waseda University, Tokyo, Japan}
\email{fujimori@phys.s.u-tokyo.ac.jp}

\title
{
Large Orbital Magnetic Moment and Strong Perpendicular Magnetic Anisotropy 
in Heavily Intercalated Fe$_{x}$TiS$_2$
} 

\abbreviations{IR,NMR,UV}

\begin{document}

\begin{abstract}
Titanium disulfide TiS$_2$, which is a member of layered transition-metal dichalcogenides 
with the 1T-CdI$_2$-type crystal structure, 
is known to exhibit a wide variety of magnetism through intercalating various kinds of 
transition-metal atoms of different concentrations. 
Among them, Fe-intercalated titanium disulfide Fe$_x$TiS$_2$ is known to be ferromagnetic 
with strong perpendicular magnetic anisotropy (PMA) and large coercive fields ($H_\text{c}$). 
In order to study the microscopic origin of the magnetism of this compound, 
we have performed X-ray absorption spectroscopy (XAS) and X-ray magnetic circular dichroism (XMCD) measurements 
on single crystals of heavily intercalated Fe$_x$TiS$_2$ ($x\sim0.5$). 
The grown single crystals showed a 
strong PMA with a 
large $H_\text{c}$ of $\mu_0H_\text{c} \simeq 1.0\ \text{T}$. 
XAS and XMCD spectra showed that Fe is 
fully in the valence states of 2+ 
and that Ti is in an itinerant electronic state, 
indicating electron transfer from the intercalated Fe atoms to the host TiS$_2$ bands. 
The Fe$^{2+}$ ions were shown to have a large orbital magnetic moment of $\simeq 0.59\ \mu_\text{B}\text{/Fe}$, 
to which, combined with 
the spin-orbit interaction and 
the trigonal crystal field,  we attribute the strong magnetic anisotropy of Fe$_x$TiS$_2$. 

\end{abstract}

\section{Introduction}

Ferromagnetism in systems where the distance between $3d$ transition-metal atoms is unusually long has attracted strong interest of researchers. Typical systems are diluted ferromagnetic semiconductors such as Ga$_{1-x}$Mn$_x$As 
\cite{Mahadevan} and double perovskites such as 
Sr$_2$FeMoO$_6$
\cite{Sarma}, where hybridization between the relatively localized $3d$ orbitals 
of the transition-metal atoms and the itinerant, more extended orbitals plays the crucial role. 

Recently, there has been keen interest in transition-metal dichalcogenides (TMDs) %
as a family of two-dimensional materials 
due to their novel physical properties \cite{Review1,Review2,Review3,BPal}. %
TMDs have stacked crystal structures 
in which two-dimensional AX$_2$ layers 
consisting of triangular lattices of the transition-metal atoms A 
sandwiched by the chalcogen atoms X 
are coupled with each other through weak van der Waals forces. 
In these compounds, 
one can intercalate other $3d$ transition-metal atoms M 
into the van der Waals gaps. 
The intercalated compounds, M$_x$AX$_2$, are known to exhibit 
a wide variety of structural, magnetic, and transport properties, including paramagnetic, ferromagnetic (FM), antiferromagnetic (AFM), and spin-glass behaviors and magneto-transport, 
depending on the host compound, the intercalated atom M, 
and its concentration $x$ \cite{Parkin_PhilMag1980, Parkin_PhilMag1980_II, InoueJMMM85, NegishiJMMM1987}. 
Titanium disulfide TiS$_2$ is one of the TMDs %
with the 1T-CdI$_2$-type layered crystal structure, 
where the Ti atom is surrounded by six S atoms octahedrally. %
Extensive studies on the 
structural \cite{Inoue_struct_JPhysChem86}, 
transport \cite{Koyano_transp_physica86, Inoue_transp_JPSJshort86}, 
and magnetic properties \cite{InoueJMMM85, NegishiJMMM1987, NegishiJMMM88} 
of M$_x$TiS$_2$ have been made so far. 
Among the transition-metal intercalated titanium disulfides, 
the Fe-intercalated compound Fe$_x$TiS$_2$ has been found to be FM 
in an exceptionally wide range of Fe concentrations $x$ $(0.2 \lesssim x \leq 1)$. 
\cite{InoueJMMM85, NegishiJMMM1987, YoshiokaJPSJ85, NegishiJMMM88, NegishiJPSJ88}. 
Moreover, the FM 
Fe$_x$TiS$_2$ has been shown to exhibit 
strong perpendicular magnetic anisotropy (PMA) %
with a large coercive field ($H_\text{c}$) of $\gtrsim 1\ \text{T}$ %
\cite{InoueJMMM85, NegishiJMMM1987, YoshiokaJPSJ85, NegishiJMMM88, NegishiJPSJ88}. 
It has also been reported that this FM 
state exhibits spin-glass-like behaviors, called the ``cluster-glass'' state, 
observed by ac susceptibility measurements \cite{InoueJMMM85,NegishiJPSJ88}, 
magnetotransport measurements \cite{Choe_PRB19}, 
and neutron scattering \cite{Kuroiwa_neut_2000}. 
Effects of the ordering of the intercalated Fe atoms, 
observed by 
transmission electron microscopy \cite{FexTiS2_TEM_JPSJ2020, Choe_PRB19}, 
are also thought to be important 
in order to explain the $x$ dependence of the magnetic phases. 

In an attempt to understand the origin of the diverse magnetic properties of these intercalated compounds M$_x$TiS$_2$ 
from the viewpoint of electronic structures, 
first-principles calculations 
\cite{Suzuki_bandcalc_JMMM87,Suzuki_bandcalc_JPSJ89,Martinez_XPSSTMcalc_JESRP02} 
and photoemission spectroscopy 
\cite{Ueda_UPS_SSC86, Ueda_UPS_JPSJ87, Fujimori_XPS_PRB88, Suga_UPSXAS_2000, FexTiS2_YamasakiSuga, Martinez_XPSSTMcalc_JESRP02,Suga_NixTiS2_ARPES_NJPhys2015}
studies have been performed so far. 
The importance of hybridization between the guest transition-metal atoms and the host TiS$_2$ 
has been pointed out based on the first-principles calculations  \cite{Suzuki_bandcalc_JMMM87,Suzuki_bandcalc_JPSJ89,Martinez_XPSSTMcalc_JESRP02}. 
Magnetic coupling between the guest transition-metal atoms thus deduced is mapped 
onto the Ruderman-Kittel-Kasuya-Yosida (RKKY) model, 
and is used to explain the complex magnetic phases as a function of the concentration $x$
\cite{Tazuke_RKKY_JPSJ2005, Tazuke_Physica06}. 
As for the magnetism of Fe$_x$TiS$_2$, X-ray absorption spectroscopy (XAS) %
and X-ray magnetic circular dichroism (XMCD) measurements %
for $x=0\text{--}0.33$ have also been performed \cite{FexTiS2_YamasakiSuga}. 
Systematic changes in the spectral line shapes as functions of $x$ for both the Fe and the Ti absorption edges 
have been observed \cite{FexTiS2_YamasakiSuga}. 
The presence of a large orbital magnetic moment of Fe observed by XMCD has also been pointed out \cite{FexTiS2_YamasakiSuga}. 
However, the relationship between the observed large orbital magnetic moment and the strong PMA has not been discussed in sufficient detail. In addition, the magnetism of Fe$_x$TiS$_2$ at higher $x$'s is more interesting because it has higher FM transition temperatures and coercive fields than that at lower $x$'s \cite{InoueJMMM85, NegishiJMMM1987}. Therefore, spectroscopic studies on heavily intercalated Fe$_x$TiS$_2$ crystals have been desired. 

In the present study, we have performed XAS and XMCD experiments 
on single crystals of heavily intercalated Fe$_x$TiS$_2$ ($x\sim0.5$). 
The grown single crystals showed a strong PMA with a
large $H_\text{c}$ of $\mu_0H_\text{c} \simeq 1.0\ \text{T}$. 
XAS and XMCD spectra 
indicate electron transfer from the intercalated Fe atoms to the TiS$_2$ host. 
The Fe$^{2+}$ ions are shown to have a large orbital magnetic moment of $\simeq 0.59 \pm 0.08\ \mu_\text{B}\text{/Fe}$, 
which would be associated with the strong magnetic anisotropy of Fe$_x$TiS$_2$.

\section{Methods}
Fe$_x$TiS$_2$ single crystals with $x \sim 0.5$ were grown by the typical chemical vapor deposition 
method. 
First, stoichiometric mixed powder with an I$_2$ transport agent in evacuated quartz tubes was heated at 500 K for 24 h to prevent the explosion of the ampule. Then, the sealed ampule was heated at 1173 K for 2 weeks with a temperature difference of 50 K 
between the both ends of the ampule. 
Detailed methods for the sample preparation are described in Ref.\ \citenum{SampleGrowth}. 
The crystallinity of the sample was checked by X-ray diffraction (XRD) 
using the X'Pert Pro diffractometer from PANalytical. 
The conventional $\theta$-$2\theta$ method in the specular geometry was adopted. 
The X-ray source was the Cu $K\alpha$ line (wavelength $\lambda = 0.15418\ \text{nm}$) 
and the operation voltage and current of the X-ray tube were 40 kV and 30 mA, respectively. 
The stoichiometry of the crystals was checked by energy dispersive X-ray (EDX) spectroscopy 
equipped with a scanning electron microscope (SEM). 
The magnetic properties of the grown crystals were measured using the 
superconducting quantum interference device 
magnetometry. 

\begin{figure}
  \includegraphics[width=8cm]{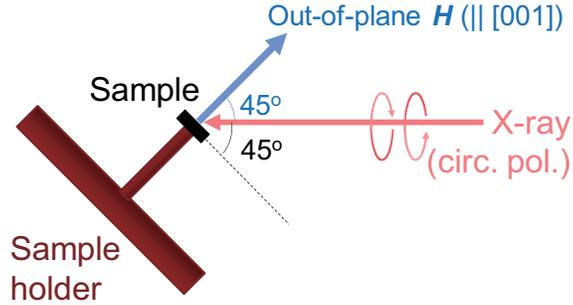}
  \caption{
  Experimental geometry of the X-ray magnetic circular dichroism (XMCD) measurements. 
  }
  \label{geometry}
\end{figure}

The XAS and XMCD spectra %
at the Fe $L_{2,3}$ ($2p \to 3d$) and Ti $L_{2,3}$ edges were measured 
using a vector-magnet XMCD measurement apparatus \cite{vector_Furuse, AngleDep_shibata}
installed at beamline BL-16A of Photon Factory, High Energy Accelerator Research Organization (KEK-PF). 
The samples were cleaved \textit{in situ} prior to measurements. 
Figure \ref{geometry}
schematically describes the experimental geometry of the XMCD measurements. 
The vacuum was maintained at $\sim 1\times 10^{-7}$ Pa during the entirety of the measurements. 
The 
samples were cooled down to 
$T=30\ $K 
without a magnetic field. 
The magnetic field was then applied parallel to the $c$-axis of the samples 
with the X-ray incident angle of 45$^\circ$. 
The maximum magnetic field was 1 T. 
XAS signals were collected in the total electron yield 
mode.

From the measured XAS and XMCD spectra, the orbital ($M_\text{orb}$) and spin ($M_\text{spin}$) magnetic moments of Fe were deduced using the XMCD sum rules \cite{SpinSum, OrbSum}.
The explicit forms of the XMCD sume rules are given as follows: \cite{SpinSum, OrbSum}
\begin{eqnarray}
M_\text{orb}&=&-\frac{4}{3}\frac{\int_{L_3+L_2}\left(\mu_{+}(E)-\mu_{-}(E)\right) dE}{\int_{L_3+L_2}\left(\mu_{+}(E)+\mu_{-}(E)\right) dE}(10-N_d), \\
M_\text{spin}&=&-\frac{2\int_{L_3}\left(\mu_{+}(E)-\mu_{-}(E)\right) dE - 4\int_{L_2}\left(\mu_{+}(E)-\mu_{-}(E)\right) dE}{\int_{L_3+L_2}\left(\mu_{+}(E)+\mu_{-}(E)\right) dE}(10-N_d), \label{spinsumeq}
\end{eqnarray}
where $\mu_{+}(E)$ and $\mu_{-}(E)$ are the XAS spectra for the right- and left-circular polarizations, respectively, 
$L_3$ ($L_2$) are the absorption peaks corresponding to the Fe $2p_{3/2} \to 3d$ ($2p_{1/2} \to 3d$) transition, 
$E$ is the photon energy, $N_d$ is the number of electrons in the transition-metal 3$d$ band, 
and $M_\text{orb}$ and $M_\text{spin}$ are given in the unit of measure of $\mu_\text{B}/\text{atom}$.
$N_d=6$, which is the nominal number of electrons in an Fe$^{2+}$ ion, was assumed in the present study. 
In the original references \cite{SpinSum, OrbSum}, 
the denominator is the sum of three XAS spectra, 
$\left(\mu_{+}(E)+\mu_{-}(E)+\mu_0(E)\right)$ 
(where $\mu_0(E)$ is the XAS spectra for the linear polarization along the X-ray incident direction), 
which were approximated by $(3/2)\left(\mu_{+}(E)+\mu_{-}(E)\right)$ here. 
In order to apply the sum rules to measured spectra, 
a continuum background was subtracted from the raw XAS spectra as follows: 
First, a polygonal line bent at the $L_3$ XAS peak position was subtracted 
so that the pre-edge and the post-edge regions became horizontal. 
Then, a smoothed two-step background was subtracted, which is composed of two arctangent functions 
with relative heights of $2:1$ centered at the peak positions of the $L_3$ and $L_2$ edges. 
We note that the arbitrariness of these background subtraction procedures causes systematic errors of 
$\sim \pm (10$--$15)\%$ in the integrated XAS intensities.
In Eq.\ (\ref{spinsumeq}), the spectra have been divided into the $L_3$- and $L_2$-edge regions at $E=718\ \text{eV}$. 
It is known that, when using the sum rules, 
the division of the XAS and XMCD spectra into the $L_3$ and $L_2$ edges at a certain energy 
may result in the underestimation of $M_\text{spin}$ \cite{Teramura, Piamonteze}. 
The presence of an additional term called the magnetic dipole term $M_T$ \cite{SpinSum, TXMCD_Stohr, TXMCD_Durr}, 
which has been omitted in Eq.\ (\ref{spinsumeq}), 
may also cause some systematic errors in $M_\text{spin}$. 
According to Ref.\ \citenum{Piamonteze}, the magnitude of the systematic errors of $M_\text{spin}$ 
due to the incomplete $L_3$-$L_2$ separation and the magnetic dipole term $M_T$ 
are estimated to be about $-10$\% and $\pm 5$\%, respectively, in the case of an Fe$^{2+}$ ion. 
Hereafter, only the statistical errors of $M_\text{orb}$ and $M_\text{spin}$ are presented. 

The spectral line shapes of the XAS and XMCD spectra were analyzed 
based on the configuration-interaction (CI) cluster model \cite{TanakaCluster} 
using the XTLS 8.5 package \cite{TanakaCluster}. 
We assumed an 
[FeS$_6$]$^{10-}$
cluster (i.e., divalent Fe) with $D_{3d}$ symmetry, 
i.e., the FeS$_6$ octahedron is slightly elongated or shrunk along a trigonal axis. 
The following parameters were adjusted in order to obtain the best-fit spectra: 
$U_{dd}$ (Coulomb energy between two Fe $3d$ valence electrons), 
$U_{dc}$ (Coulomb energy between the Fe $3d$ electron and the Fe $2p$ core electron), 
$\Delta$ (charge-transfer energy), 
$(pd\sigma)$ (Slater-Koster parameters), 
$10Dq$ (octahedral crystal field), 
$D_\text{trg}$ (trigonal crystal field), and 
$H_\text{mol}$ (molecular field). 
Spin-orbit interaction (SOI) %
and the Slater integrals were estimated from atomic Hartree-Fock calculations. 
For the Slater integrals, 80\% of the values deduced from the atomic Hartree-Fock calculation were used. 
In the CI calculation, the ground state was assumed to be a linear combination of 
the $3d^6$, $3d^7\underline{L}$, and $3d^8\underline{L}^2$ configurations, 
where $\underline{L}$ is a ligand hole. 

\section{Results and Discussion}

\subsection{Sample Characterization}

\begin{figure}
  \includegraphics[width=10cm]{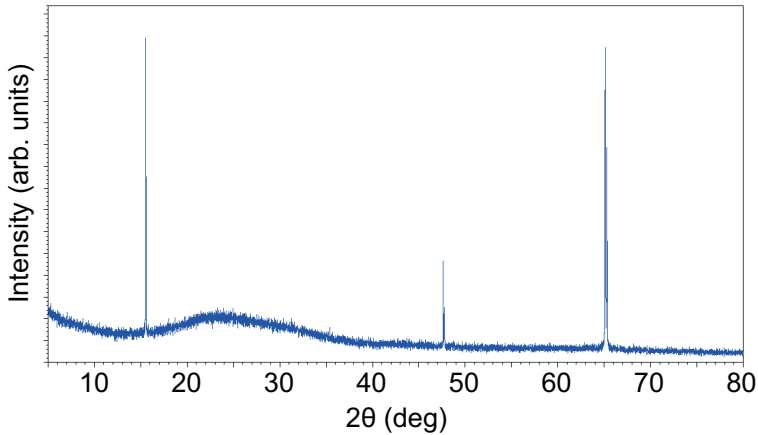} 
  \caption{
  X-ray diffraction (XRD) profile along the out-of-plane [001] direction of an Fe$_{x}$TiS$_2$ ($x\sim0.5$) single crystal.  
  }
  \label{XRD}
\end{figure}

Figure \ref{XRD} shows the XRD profile of an Fe$_{x}$TiS$_2$ ($x\sim0.5$) single crystal. 
Sharp Bragg peaks along the [001] direction are clearly observed, 
confirming the layered structure of the grown crystal. 
From the obtained XRD profile, the out-of-plane lattice parameter $c$ has been estimated to be 0.572 nm, 
close to the value $c=0.574\ \text{nm}$ obtained in the previous study 
for Fe$_{x}$TiS$_2$ with $x= 0.5$ \cite{Inoue_struct_JPhysChem86}. 

\begin{figure}
  \includegraphics[width=12cm]{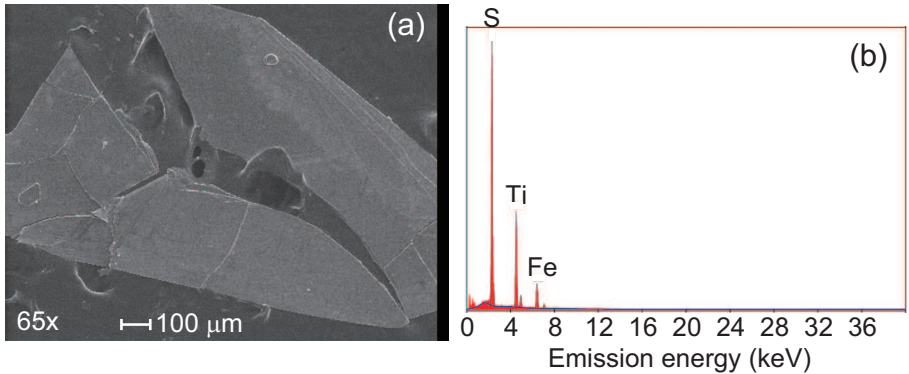}
  \caption{
  Energy dispersive X-ray spectroscopy (EDX) analysis of an Fe$_{x}$TiS$_2$ single crystal. 
  (a) Scanning electron microscope (SEM) image of the Fe$_{x}$TiS$_2$ single crystal on a carbon tape. 
  (b) EDX spectrum. 
  }
  \label{EDX}
\end{figure}

Figure \ref{EDX}(a) shows the SEM image of the Fe$_{x}$TiS$_2$ single crystal on a carbon tape. 
The typical sample diameter is 0.5--1 mm. 
Regions with flat surfaces can be seen 
with typical sizes of a few hundred micrometers squared. 
The EDX spectrum of the crystal is shown in Fig.\ \ref{EDX}(b). 
From the spectral intensities, the atomic ratio of each element 
has been estimated 
to be $\text{Fe}:\text{Ti}:\text{S}=12.7:28.5:58.8$, 
which corresponds to the chemical formula Fe$_{0.432}$Ti$_{0.968}$S$_2$.
This confirms that more Fe atoms are intercalated in the present Fe$_{x}$TiS$_2$ crystal 
than those in the previous XMCD study ($x\leq 0.33$), \cite{FexTiS2_YamasakiSuga}
although the Fe concentration is slightly lower than the nominal value $(x=0.5)$. 

\begin{figure}
  \includegraphics[width=8cm]{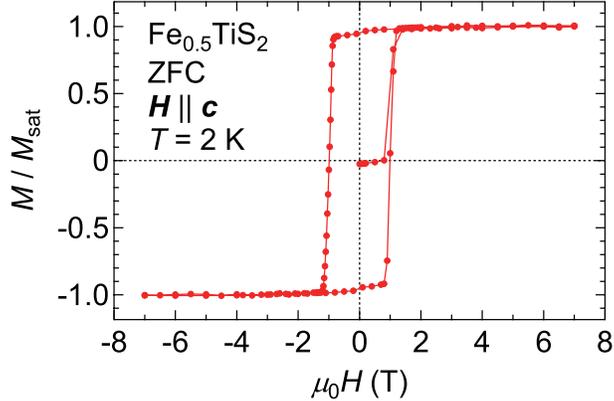}
  \caption{Normalized magnetization curve of the Fe$_{0.5}$TiS$_2$ single crystal. 
  }
  \label{MH}
\end{figure}

From the temperature dependence of the magnetization, 
the Curie temperature of the crystal has been estimated to be $\sim 70\ \text{K}$. 
Figure \ref{MH} shows the 
zero-field cooled (ZFC) 
magnetization curve 
of the Fe$_{x}$TiS$_2$ $(x\sim 0.5)$ single crystal 
measured at 2 K 
with an applied magnetic field $\bm{H}$ parallel to the $c$-axis (trigonal axis) 
up to $\mu_0 |\bm{H}|= 7\ $T. 
In the initial magnetization process, a sudden jump in the magnetization $M$ is observed 
around $\mu_0 H \sim 1\ \text{T}$, suggesting 
a transition from the AFM to FM states. %
After magnetic saturation, 
a 
clear rectangular hysteresis loop is observed 
with a coercive field $H_\text{c}$ of $\mu_0 H_\text{c} \sim 1\ \text{T}$. 
The obtained $H_\text{c}$ is similar to those of polycrystalline Fe$_x$TiS$_2$ with ($x=0.33\text{--}0.45$) 
and single-crystalline Fe$_{0.33}$TiS$_2$ 
reported in the previous study \cite{NegishiJMMM1987}. 
The large hysteresis loop suggests that the crystal exhibits a strong PMA 
with the magnetic easy axis along the $c$-axis. 
We note that quantitative estimate of the saturation magnetization $M_\text{sat}$ from the raw magnetization curve 
is difficult due to a large uncertainty of the sample mass. 
We have, therefore, plotted the magnetization curve normalized by $M_\text{sat}$ in Fig.\ \ref{MH}. 
We shall deduce the values of $M_\text{sat}$ and the uniaxial magnetic anisotropy energy (MAE) $K_\text{u}$ 
by the XMCD measurements afterward. 

\subsection{X-ray Magnetic Circular Dichroism Spectroscopy}

\begin{figure}
  \includegraphics[width=16cm]{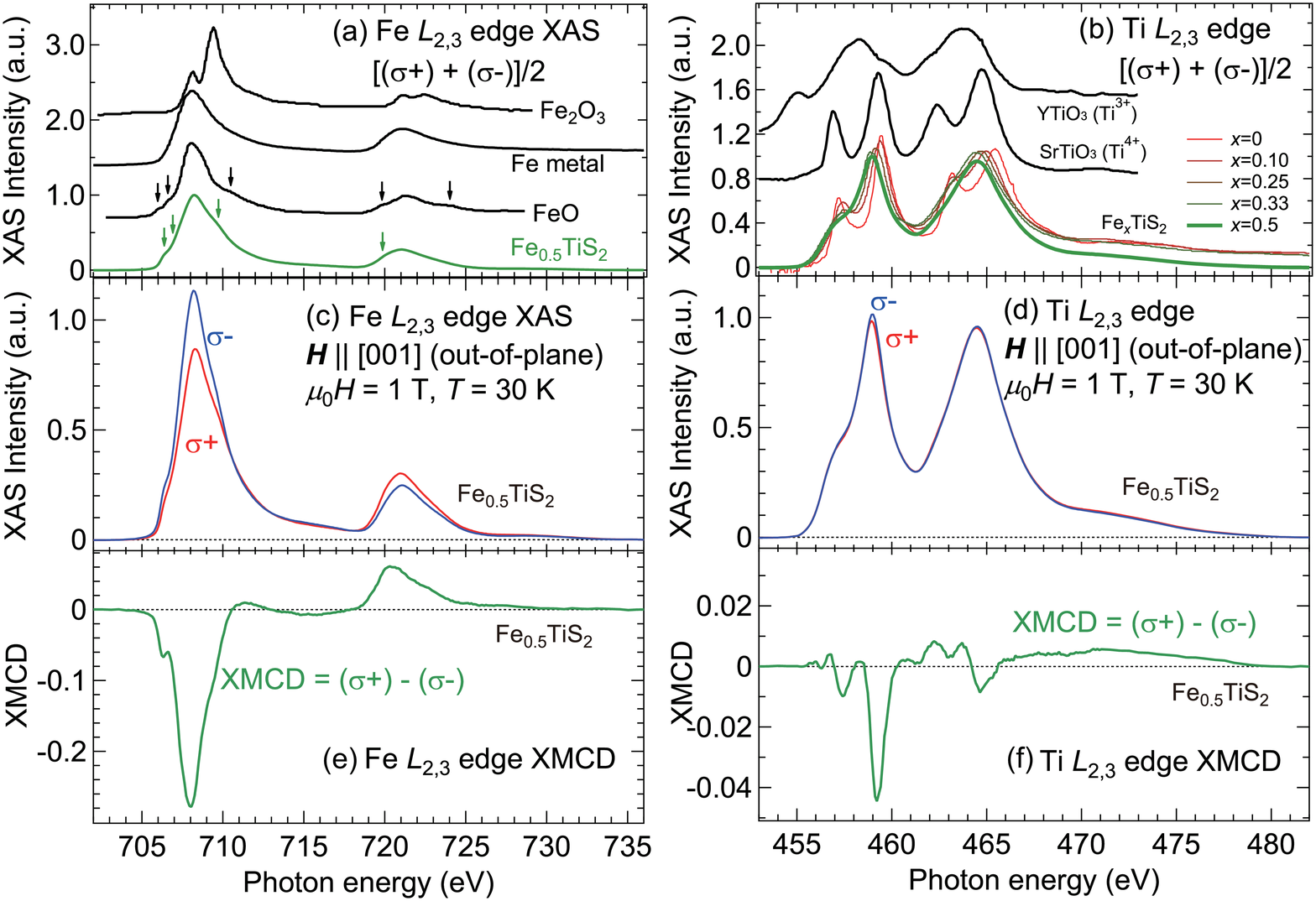}
  \caption{
  X-ray absorption spectroscopy (XAS) and XMCD spectra. 
  (a,b) Helicity-averaged XAS spectrum of Fe$_{0.5}$TiS$_2$ at the Fe $L_{2,3}$ ($2p_{1/2, 3/2} \to 3d$) (a) 
  and Ti $L_{2,3}$ (b) absorption edges. 
  The reference XAS spectra of iron 
oxides, metallic iron, 
%
and titanium oxides 
  are taken from Refs.\ \citenum{FeXASref_Regan} 
, \citenum{CTChen}, 
%
and \citenum{TiXASref_Cao}, respectively. 
  The Ti XAS spectra of TiS$_2$ and Fe$_{x}$TiS$_2$ ($0\leq x \leq 0.33$) are taken from 
  Refs.\ \citenum{Kimura_TiS2} and \citenum{FexTiS2_YamasakiSuga}, respectively. 
Arrows in Panel (a) indicate the multiplet structures of Fe$^{2+}$, which are absent in the metallic Fe. 
%
  (c,d) Helicity-dependent
  XAS spectra of the Fe$_{0.5}$TiS$_2$ single crystal at the Fe $L_{2,3}$ (c) and Ti $L_{2,3}$ (d) edges. 
  (e,f) XMCD spectra at the Fe $L_{2,3}$ (e) and Ti $L_{2,3}$ (f) edges. 
 The spectra were measured at the temperature ($T$) of 30 K 
 under the magnetic field ($\mu_0 H$) of 1 T applied parallel to the $c$-axis ([001] axis). 
  The spectra have been normalized so that the peak height of the helicity-averaged XAS spectra at the $L_3$ edge is equal to unity. 
   In Panels (a)--(d), two-step backgrounds has been subtracted from the raw XAS spectra. 
  }
  \label{XASXMCD}
\end{figure}

Figure \ref{XASXMCD}(a) 
shows the helicity-averaged XAS spectra 
of the Fe$_{x}$TiS$_2$ $(x\sim 0.5)$ single crystal
at the Fe $L_{2,3}$ ($2p_{1/2, 3/2} \to 3d$) 
absorption edges. 
The XAS spectra of divalent and trivalent iron oxides 
and metallic iron 
%
are also shown as references 
\cite{FeXASref_Regan, CTChen}. 
%
The overall spectral line shape of Fe$_{0.5}$TiS$_2$ is similar to that of FeO. 
Especially, the small shoulder around $h\nu = 706.5\ \text{eV}$ is characteristic of divalent Fe, 
showing that the iron atom is essentially in the Fe$^{2+}$ valence state. 
Reflecting the large hybridization strength of sulfides and high electrical conductivity compared to oxides, 
the XAS spectrum of Fe$_{0.5}$TiS$_2$ shows broader absorption peaks 
and more extended high-energy tails than that of FeO. 
One may think that the XAS spectrum of Fe$_{0.5}$TiS$_2$ is also similar to that of the Fe metal. 
As indicated in Fig.\ \ref{XASXMCD}(a) by arrows, however, 
the XAS spectra of Fe$_{0.5}$TiS$_2$ and FeO have several multiplet structures in common, 
while they are absent in the XAS spectrum of metallic Fe reflecting its itinerant electronic structures. 
This indicates that the amount of possible metallic iron clusters, if any, is small in the present Fe$_{0.5}$TiS$_2$ crystal. 
The difference of the peak positions of the multiplet structures between 
Fe$_{0.5}$TiS$_2$ and FeO is probably due to the different hybridization strength and crystal-field splitting 
between oxides and sulfides. 

In Fig.\ \ref{XASXMCD}(b), 
the helicity-averaged XAS spectra 
of the Fe$_{x}$TiS$_2$ $(x\sim 0.5)$ single crystal
at the Ti $L_{2,3}$ absorption edges 
are compared with those of Fe$_x$TiS$_2$ ($0\leq x\leq 0.33$) \cite{FexTiS2_YamasakiSuga, Kimura_TiS2} 
and titanium oxides \cite{TiXASref_Cao}. 
The spectral line shape of the present Fe$_{0.5}$TiS$_2$ crystal 
is close to those of Fe$_x$TiS$_2$ with $x \leq 0.33$ \cite{FexTiS2_YamasakiSuga}. 
One can see that, as $x$ increases, 
the absorption peaks become broader and the energy splitting between the two principal peaks 
in each of the $L_3$ and $L_2$ edges become systematically smaller. 
The shoulder structures around $h\nu = 457$ and $463\ \text{eV}$ 
in the present XAS spectra are more diffuse than those for $x=0.33$, 
confirming that a larger amount of Fe atoms are intercalated in the present sample than in the previous ones. 
These spectral changes with increasing $x$ are considered to be due to electron doping 
from the intercalated Fe atoms into the TiS$_2$ host. 
As can be seen from the figure, the spectral line shapes of Fe$_x$TiS$_2$ with $x\neq 0$ cannot be described 
as a superposition of the Ti$^{4+}$ and Ti$^{3+}$ valence states. 
This implies that the electrons doped from Fe into TiS$_2$ 
do not lead to a mixed-valence state consisting of the Ti$^{4+}$ and Ti$^{3+}$ ionic states, but 
have rather itinerant character
. 

Figures \ref{XASXMCD}(c)--\ref{XASXMCD}(f) show the helicity-dependent XAS spectra 
and the XMCD spectra at the Fe $L_{2,3}$ and Ti $L_{2,3}$ edges. 
The XMCD spectra in Figs.\ \ref{XASXMCD}(e) and \ref{XASXMCD}(f) clearly show that 
not only Fe but also Ti exhibits finite XMCD signals. 
This suggests that the observed ferromagnetism 
is not due to magnetic impurities such as metallic iron or iron compounds 
but due to the intrinsic properties of the Fe$_{0.5}$TiS$_2$ crystal. 
First-principle calculations 
\cite{Suzuki_bandcalc_JMMM87, Suzuki_bandcalc_JPSJ89} 
show that the band structure of the transition-metal-intercalated TiS$_2$ 
is significantly modified from that of the nonintercalated TiS$_2$ 
through hybridization between the intercalated atoms and the host compound. 
In the case of Fe intercalation, 
the down-spin Fe 3$d$ states hybridize with the Ti 3$d$ states 
to form spin-polarized bands near the Fermi level \cite{Suzuki_bandcalc_JMMM87}. 
As a result, Ti exhibits small but finite spin polarization
\cite{Suzuki_bandcalc_JMMM87}. 
The present XMCD result is qualitatively consistent with this calculation. 

\begin{figure}
  \includegraphics[width=8cm]{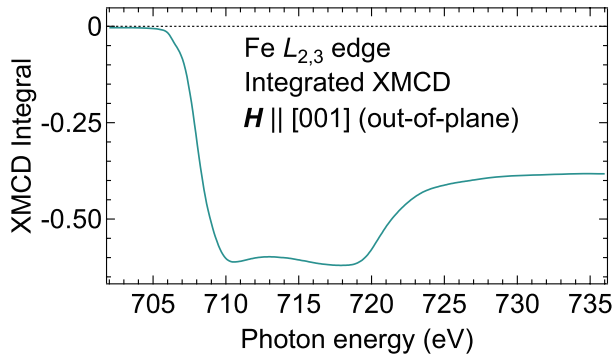}
  \caption{Integrated XMCD spectra of the Fe$_{0.5}$TiS$_2$ single crystal at the Fe $L_{2,3}$ edge.}
  \label{XMCDinteg}
\end{figure}

Using the XMCD sum rules 
\cite{SpinSum, OrbSum}, 
we have deduced the spin ($M_\text{spin}$) and orbital magnetic moments ($M_\text{orb}$) of Fe. 
We note that the $M_\text{spin}$ of Ti cannot be deduced from the spin sum rule 
due to the too small spin-orbit splitting of the Ti $2p$ core levels, 
making the decomposition of the XMCD spectrum into the $L_2$ and $L_3$ components 
\textcolor{red}{impossible}
\cite{SpinSum, Teramura, Piamonteze}. 
We also note that the $M_\text{orb}$ of Ti, which is obtained by integrating the entire XMCD spectrum, 
was below the detection limit ($\sim 0.1\ \mu_\text{B}\text{/atom}$) 
of the present experiment. 
Considering that the magnetization is 45$^\circ$ off from the X-ray incident direction, 
the obtained magnetic moments have been divided by $\cos 45^\circ = 1/\sqrt{2}$ 
in order to correct this effect. 
As shown in Fig.\ \ref{XMCDinteg}, 
the integral of the Fe XMCD spectrum has a large negative end value, 
suggesting that Fe has a large $M_\text{orb}$ parallel to the spin magnetic moment. 
The results of the sum rule yield $M_\text{spin}=2.45 \pm 0.16\ \mu_\text{B}\text{/Fe}$ 
and $M_\text{orb}=0.59 \pm 0.08\ \mu_\text{B}\text{/Fe}$. 
The obtained $M_\text{spin}$ is consistent with the values 
deduced from the first-principle calculations: 
$M_\text{spin}=2.45 \mu_B/\text{Fe}$ for $x=1/3$ and $M_\text{spin}=2.99 \mu_B/\text{Fe}$ for $x=1$
\cite{Suzuki_bandcalc_JMMM87}.
The large $M_\text{orb}$ may be due to the degeneracy of the electron orbitals in the Fe$^{2+}$ valence state 
and the partially filled $t_{2g}$ level in a 
relatively weak crystal fields for the intercalated Fe atoms. %

From the $M_\text{spin}$ and $M_\text{orb}$ values deduced by using the XMCD sum rule, 
one can deduce the uniaxial magnetic anisotropy energy $K_\text{u}$ 
based on the Stoner-Wohlfarth model, \cite{SWmodel} 
in which one assumes uniaxial magnetic anisotropy 
and the coherent rotation of the magnetic domain. 
According to the model, 
the coercive field $H_\text{c}$ is equal to $2K_\text{u}/(\mu_0 M_\text{sat})$, 
where $M_\text{sat} = 3.04\pm0.18\ \mu_\text{B}\text{/Fe}$ is 
the saturation magnetization calculated as the sum of $M_\text{spin}$ and $M_\text{orb}$. 
From this relationship, the uniaxial MAE for the present Fe$_{x}$TiS$_2$ single crystal 
is estimated to be $K_\text{u} = 0.176\pm0.011\ \text{meV/Fe}$ 
or $K_\text{u} = 0.244\pm 0.015\ \text{MJ/m}^3$ 
(note that $K_\text{u}$ is defined positive for materials with PMA). 

\subsection{CI Cluster-Model Calculation}
\begin{figure}
  \includegraphics[width=16cm]{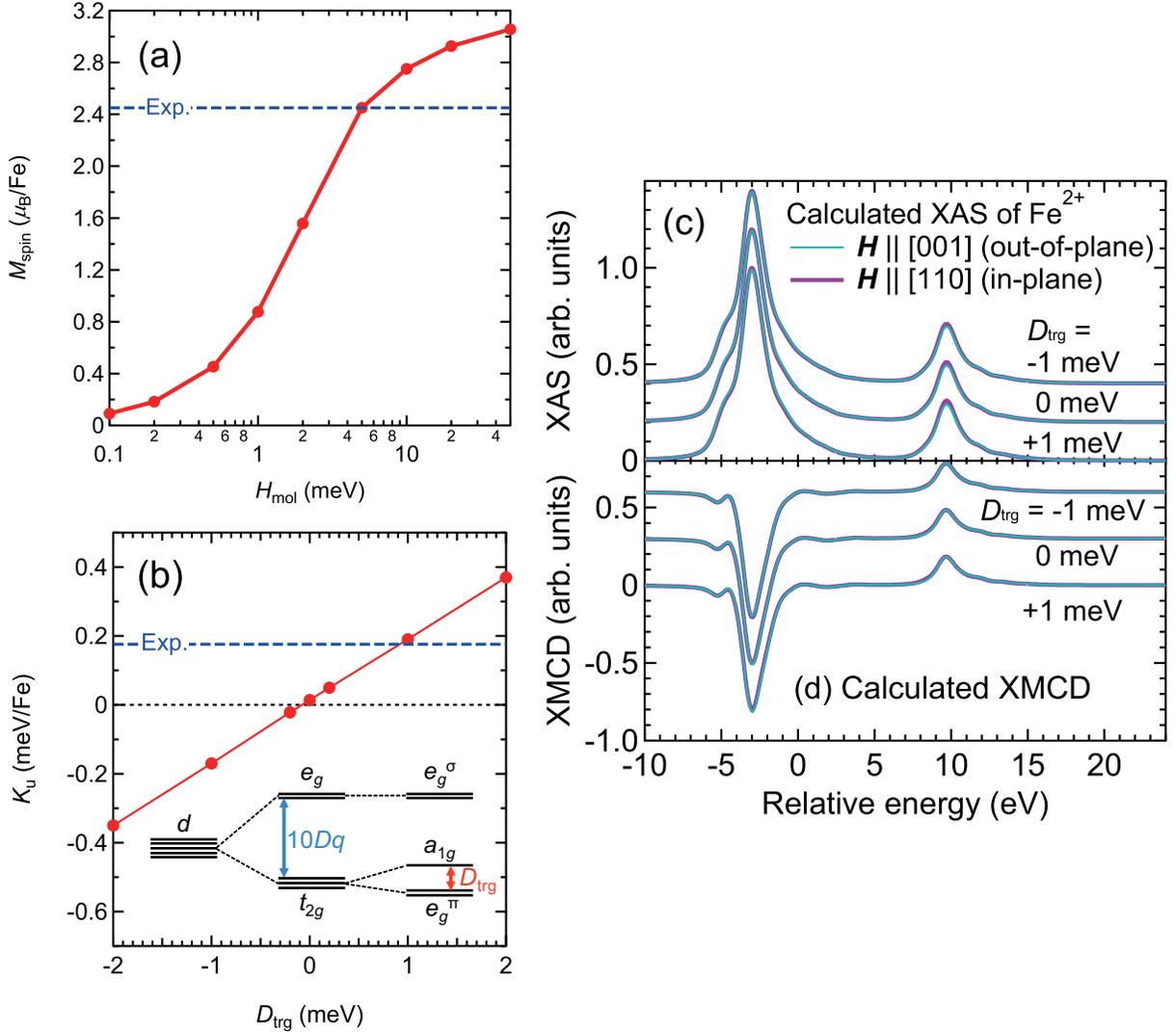}
  \caption{Results of CI cluster-model calculations at the Fe $L_{2,3}$ edge. 
  (a) Molecular field ($H_\text{mol}$) dependence of the spin magnetic moment per Fe atom ($M_\text{spin}$). 
  (b) Trigonal crystal field ($D_\text{trg}$) dependence of the magnetic anisotropy energy per Fe atom ($K_\text{u}$). 
  In panels (a) and (b), blue dashed lines represent the experimental values. 
  Inset of panel (b) depicts the crystal-field splitting in the case of $D_\text{trg}>0$. 
  (c),(d) Calculated XAS (c) and XMCD (d) spectra for $D_\text{trg}=0\ \text{meV}$ and 
  $D_\text{trg}=\pm 1\ \text{meV}$. 
  See Table \ref{ClusterPara} for the other parameter values used. }
  \label{Cluster}
\end{figure}

In order to explain the observed large $M_\text{orb}$ of Fe 
and the strong PMA of $\mu_0 H_\text{c} \simeq 1.0\ \text{T}$
and $K_\text{u} = 0.244\ \text{MJ/m}^3$, 
we calculated the XAS and XMCD spectra at the Fe $L_{2,3}$ edges 
using the CI cluster model. 
An [FeS$_6$]$^{-10}$ cluster 
with weak trigonal distortion ($D_{3d}$ symmetry) was employed. 
In the present calculation, we first optimized the parameters 
$\Delta$, $U_{dd}$, $U_{dc}$, $(pd\sigma)$, and $10Dq$ in the $O_h$ symmetry (i.e., $D_\text{trg}=0$) 
so that the calculated spectra well reproduce the experimental ones 
(for the definitions of $10Dq$ and $D_\text{trg}$, see the inset of Fig.\ \ref{Cluster}(b))
. 
The best-fit result was obtained with the parameter values listed in Table \ref{ClusterPara}. 
The obtained parameter values are comparable to 
those deduced from the photoemission spectroscopy of Fe$_{0.33}$TiS$_2$ \cite{Bouquet_PRB1991}. 
We then optimized the values of $H_\text{mol}$ and $D_\text{trg}$ 
so that they reproduce the experimental $M_\text{spin}$ and $K_\text{u}$, respectively. 
$K_\text{u}$ has been calculated as the total energy difference between two 
spin configurations, $\bm{M}_\text{spin} \parallel [001]$ and $\bm{M}_\text{spin} \parallel [110]$. 
Figure \ref{Cluster}(a) shows the $H_\text{mol}$ dependence of the calculated spin magnetic moment $M_\text{spin}$. 
We note that modifying the values of $H_\text{mol}$ only resulted in the 
change in the XMCD intensity, without any variation in the characteristic spectral line shapes (not shown here). %
By comparing the calculated $M_\text{spin}$ with the experimental value, 
$H_\text{mol}$ has been estimated to be $5 \pm 1\ \text{meV}$. 
Figure \ref{Cluster}(b) shows the $D_\text{trg}$ dependence of the calculated magnetic anisotropy energy $K_\text{u}$. 
It can be seen that the experimental $K_\text{u}$ can be reproduced by 
only a slight change in $D_\text{trg}$ of $\sim 1\ \text{meV}$. 
The calculated Fe $L_{2,3}$ XAS and XMCD spectra 
are shown in Figs.\ \ref{Cluster}(c) and \ref{Cluster}(d) 
for $D_\text{trg}=0\ \text{meV}$ and $D_\text{trg}=\pm 1 \ \text{meV}$. 
The calculated spectra well reproduce the experimental spectral features. 
It can also be seen that the spectral line shapes and the XMCD intensities are nearly insensitive to $D_\text{trg}$, 
although such a small magnitude of $D_\text{trg}$ can induce appreciable magnetic anisotropy as shown below.

\begin{table}
  \caption{Best-fit parameters for the configuration-interaction (CI) cluster-model calculations.}
  \label{ClusterPara}
  \begin{tabular}{ccccc}
    \hline
    $\Delta $ & $U_{dd}$ & $U_{dc}$ & $(pd\sigma)$ & $10Dq$ \\
    \hline
    2.5 eV & 5.0 eV & 6.3 eV & -0.8 eV & 0.7 eV \\
    \hline
  \end{tabular}
\end{table}

\subsection{Discussion: Origin of the Strong PMA}

The microscopic origin of the strong PMA of Fe-intercalated TMDs is discussed by Parkin and Friend \cite{Parkin_PhilMag1980} 
based on the atomic 
multiplet
model in an octahedral crystal field with a trigonal distortion. 
Here, we summarize their discussion and apply it to the present system. 
An Fe$^{2+}$ ion with $3d^6$ electron configuration octahedrally coordinated by six anions has the 
$^{5}T_{2g}$ ground state, 
which is triply degenerate in the orbital part ($(t_{2g})^4(e_g)^2$) and quintuply degenerate in the spin part ($S=2$). 
(Note that the multielectron state $^{5}T_{2g}$ should be distinguished from the one-electron state $t_{2g}$.)
. 
The orbital triplet is isomorphic to the substates of orbital angular momentum $L=1$. 
In order to represent this orbital triplet, an effective orbital angular momentum operator $\bm{L}^\text{eff}$ is introduced. 
Using $\bm{L}^\text{eff}$, the Hamiltonian of the system can be written as follows: 
\begin{equation}
\hat{H} = -D_\text{trg} (L_z^\text{eff})^2 - \lambda \bm{L}^\text{eff} \cdot \bm{S} 
+ \bm{S}\cdot \bm{H}_\text{mol},
\label{hamil}
\end{equation}
where the first, second, and third 
terms represent the energies of the trigonal crystal field $D_\text{trg}$ $(>0)$, SOI of the $3d$ electrons $\lambda$ $(>0)$, 
and the Zeeman effect by the molecular field $\bm{H}_\text{mol}$ (in the unit of measure of eV) originating from exchange interactions, respectively. 
Note that we have replaced the Zeeman term due to the external magnetic field $\mu_B (-\bm{L}^\text{eff} + 2\bm{S}) \cdot \mu_0 \bm{H}$, which is originally introduced by Parkin and Friend \cite{Parkin_PhilMag1980}, by that due to the molecular field 
because we consider the FM state instead of the paramagnetic one. 
They have treated the first two terms as the primary terms and the Zeeman term as perturbation 
and have shown that the energy corrections for the out-of-plane and in-plane magnetic fields 
are of the first and second orders in $H_\text{mol}$, respectively, 
resulting in the anisotropy of the $g$-factor. 
However, our cluster-model calculation shows that $\lambda\ (\simeq 50\ \text{meV}) > H_\text{mol}\ (\simeq 5\ \text{meV}) > D_\text{trg}\ (\simeq 1\ \text{meV})$. 
We, therefore, treat the role of $D_\text{trg}$ as perturbation to SOI and $H_\text{mol}$ 
in order to explain the magnetic anisotropy of FM Fe$_x$TiS$_2$. 

\begin{figure}
  \includegraphics[width=16cm]{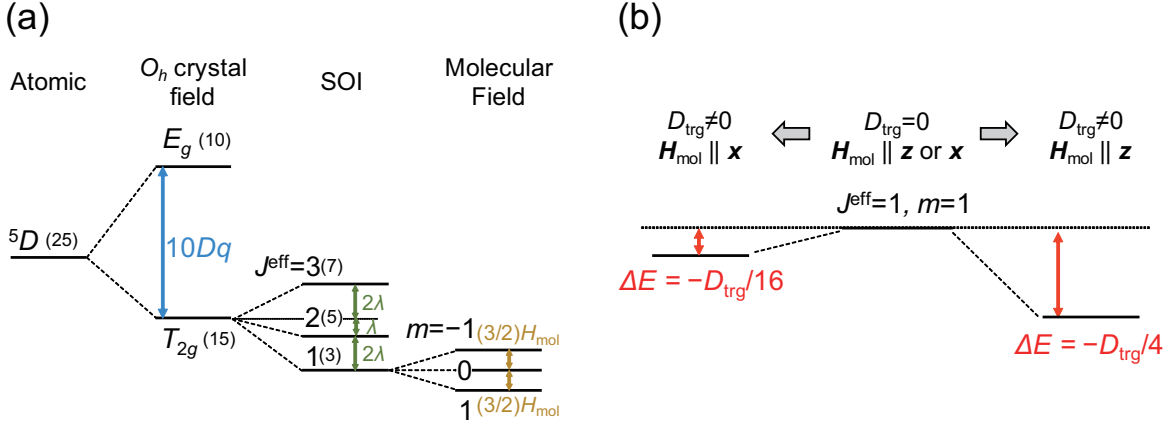}
  \caption{
  Schematic energy diagram of the Fe$^{2+}$ ion ($3d^6$ electron configuration) 
  octahedrally coordinated by six anions with small trigonal distortion. 
  (a) 
  Energy splitting due to the octahedral crystal field ($10Dq$), SOI ($\lambda$), and molecular field ($H_\text{mol}$). 
  The numbers in parentheses denote the degeneracy. 
  (b) Energy shift due to the trigonal crystal field $D_\text{trg}$ 
  in the cases of out-of-plane ($\bm{H}_\text{mol} \parallel \bm{z}$) and in-plane molecular fields ($\bm{H}_\text{mol} \parallel \bm{x}$). 
}
  \label{Ediag}
\end{figure}

Figure \ref{Ediag}(a) schematically describes the energy diagram of an Fe$^{2+}$ ion based on this model. 
The $^{5}T_{2g}$ ground state under the octahedral crystal field is split into three levels of $J^\text{eff}=1,2,3$
by SOI, where $\bm{J}^\text{eff}=\bm{L}^\text{eff}+\bm{S}$ is the effective total angular momentum. 
Since $\lambda$ is 1 order of magnitude larger than the thermal energy $k_B T\ (\sim 2.5\ \text{meV})$, 
only the $J^\text{eff}=1$ level has to be considered. 
The $J^\text{eff}=1$ 
level is then 
Zeeman-split into three sublevels of $m=1,0,-1$, where $m$ is the magnetic quantum number for $\bm{J}^\text{eff}$. 
The magnetic anisotropy energy can be deduced by calculating the difference of the perturbation energy of $D_\text{trg}$ in the cases of out-of-plane ($\bm{H}_\text{mol} \parallel \bm{z}$) and in-plane molecular fields ($\bm{H}_\text{mol} \parallel \bm{x}$). 
The ground state $\displaystyle | J^\text{eff}=1,\ m=1 \rangle$ is expressed as follows in the cases of ($\bm{H}_\text{mol} \parallel \bm{z}$) and ($\bm{H}_\text{mol} \parallel \bm{x}$): 
\begin{eqnarray*}
| J^\text{eff}=1,\ m=1 \rangle_z &=& | J^\text{eff}=1,\ J_z^\text{eff}=1 \rangle \qquad (\bm{H}_\text{mol} \parallel \bm{z}), \\
| J^\text{eff}=1,\ m=1 \rangle_x &=&
 \frac{1}{2}| J^\text{eff}=1,\ J_z^\text{eff}=1 \rangle \\
 & & +\frac{1}{\sqrt{2}}| J^\text{eff}=1,\ J_z^\text{eff}=0 \rangle \\
 & & \qquad + \frac{1}{2}| J^\text{eff}=1,\ J_z^\text{eff}=-1 \rangle 
 \qquad (\bm{H}_\text{mol} \parallel \bm{x}), 
\end{eqnarray*}
where the $\displaystyle | J^\text{eff}=1,\ J_z^\text{eff} \rangle$ terms can be written using the eigenstates of $L_z^\text{eff}$ and $S_z$ (denoted as $|L_z,\ S_z\rangle$) as 
\begin{eqnarray*}
	|J^\text{eff}=1,\ J_z^\text{eff}=1 \rangle
	 &=& \sqrt{\frac{6}{10}}|-1,2\rangle - \sqrt{\frac{3}{10}}|0,1\rangle+ \sqrt{\frac{1}{10}}|1,0\rangle, \\
	|J^\text{eff}=1,\ J_z^\text{eff}=0 \rangle
	 &=& \sqrt{\frac{3}{10}}|-1,1\rangle - \frac{2}{\sqrt{10}}|0,0\rangle+ \sqrt{\frac{3}{10}}|1,-1\rangle, \\
	|J^\text{eff}=1,\ J_z^\text{eff}=-1 \rangle
	 &=& \sqrt{\frac{1}{10}}|-1,0\rangle - \sqrt{\frac{3}{10}}|0,-1\rangle+ \sqrt{\frac{6}{10}}|1,-2\rangle, 
\end{eqnarray*}
Thus, the perturbation energies due to the trigonal crystal field 
$\displaystyle \langle -D_\text{trg} \left( L_z^\text{eff} \right) ^2 \rangle$ 
for the out-of-plane and in-plane fields can be calculated as $-D_\text{trg}/4$ and $-D_\text{trg}/16$, respectively, to the first order of 
$D_\text{trg}$. 
This energy shift is schematically described in Fig.\ \ref{Ediag}(b). 
This energy difference can be regarded as the origin of the PMA in the present system. 
We stress that this is a first-order perturbation effect, 
which is the reason why such a small magnitude of $D_\text{trg}$ can give rise to the considerable magnitude of the magnetic anisotropy energy. 
The present result is consistent with the $D_\text{trg}$ dependence of $K_\text{u}$ shown in Fig.\ \ref{Cluster}(b), in which $K_\text{u}$ is nearly proportional to $D_\text{trg}$ when the crystal field is weak. 

The present XMCD results show that there exists a large unquenched orbital moment of the intercalated Fe atom ($\simeq 0.59 \mu_\text{B}\text{/Fe}$) in Fe$_{0.5}$TiS$_2$, which is considered to originate from the above-mentioned effective orbital angular momentum in the $^{5}T_{2g}$ state. 
The ratio of the orbital to spin angular momentum ($\langle L_z \rangle / \langle S_z \rangle = 2M_\text{orb}/M_\text{spin}$) is calculated to be $0.48 \pm 0.07$. 
According to the previous XMCD measurements for Fe$_{x}$TiS$_{2}$ with $x \leq 0.33$ \cite{FexTiS2_YamasakiSuga}, 
the values of $\langle L_z \rangle / \langle S_z \rangle$ are deduced to be $0.49,\ 0.48,$ and $0.25$ for $x=0.10,\ 0.25,$ and $0.33$, respectively. 
The $\langle L_z \rangle / \langle S_z \rangle$ value obtained for $x=0.5$ in the present study is 
out of this decreasing tendency with increasing $x$. 
Although the mechanism for this $x$ dependence of $\langle L_z \rangle / \langle S_z \rangle$ is not clear at present, 
hybridization effects between guest Fe and host TiS$_{2}$, 
which have not been taken into account in the above discussion, 
might be relevant. 
We note that similar $x$ dependence has also been observed in the strength of the coercive field $H_\text{c}$ of Fe$_x$TiS$_2$ 
\cite{InoueJMMM85, NegishiJMMM1987}. 
We expect that the complex $x$ dependence of $H_\text{c}$ can be explained by the magnitude of the unquenched orbital magnetic moment of the intercalated Fe$^{2+}$ ions. 

\section{Conclusion}
We have performed XAS and XMCD experiments 
on single crystals of heavily intercalated Fe$_x$TiS$_2$ ($x\sim0.5$) 
in order to discuss the microscopic origin of the strong PMA and large $H_\text{c}$ of this compound. 
We have confirmed that the grown single crystal showed a large $H_\text{c}$ of $\mu_0H_\text{c} \simeq 1.0\ \text{T}$ and 
strong PMA %
with a magnetic anisotropy energy $K_\text{u}$ of 
$0.244\ \text{MJ/m}^3$. 
From the XMCD result, the Fe$^{2+}$ ions are shown to have a large orbital magnetic moment of $\simeq 0.59 \pm 0.08\ \mu_\text{B}\text{/Fe}$. 
The strong magnetic anisotropy of Fe$_x$TiS$_2$ is attributed to the large orbital moment combined with 
the spin-orbit interaction and 
the trigonal crystal field.

\begin{acknowledgement}

A.F. and S.W.C. would like to thank D.\ D.\ Sarma for long-standing collaboration on a variety of interesting $3d$ transition-metal compounds and for stimulating interaction during the last many decades.  
We thank Kenta Amemiya and Masako Sakamaki 
for valuable technical support at KEK-PF. 
This work was supported by Grants-in-Aid for Scientific Research from 
the JSPS (15H02109, 15K17696, 19K03741, and 20K14416). 
J.W.K. and S.W.C. were supported by the National Science Foundation (NSF) under Grant No. DMR-1629059. 
C.J.W. and S.W.C. partially were supported by the National Research Foundation of Korea (NRF) funded by the Ministry of Science and ICT(No. 2016K1A4A4A01922028 and No. 2020M3H4A2084417). 
Y.X.W.\ acknowledges support from 
Advanced Leading Graduate Course for Photon Science (ALPS) 
at the University of Tokyo. 
A.F.\ is an adjunct member of Center for Spintronics Research Network (CSRN), 
the University of Tokyo, under Spintronics Research Network of Japan (Spin-RNJ). 
The experiment was done under the approval of the Photon Factory 
Program Advisory Committee 
(Proposal No.\ 2016S2-005, No.\ 2016G066, and No.\ 2019G622). 

\noindent The authors declare no competing financial interest. 

\end{acknowledgement}

\bibliography{ref}

\providecommand{\latin}[1]{#1}
\makeatletter
\providecommand{\doi}
  {\begingroup\let\do\@makeother\dospecials
  \catcode`\{=1 \catcode`\}=2 \doi@aux}
\providecommand{\doi@aux}[1]{\endgroup\texttt{#1}}
\makeatother
\providecommand*\mcitethebibliography{\thebibliography}
\csname @ifundefined\endcsname{endmcitethebibliography}
  {\let\endmcitethebibliography\endthebibliography}{}
\begin{mcitethebibliography}{47}
\providecommand*\natexlab[1]{#1}
\providecommand*\mciteSetBstSublistMode[1]{}
\providecommand*\mciteSetBstMaxWidthForm[2]{}
\providecommand*\mciteBstWouldAddEndPuncttrue
  {\def\EndOfBibitem{\unskip.}}
\providecommand*\mciteBstWouldAddEndPunctfalse
  {\let\EndOfBibitem\relax}
\providecommand*\mciteSetBstMidEndSepPunct[3]{}
\providecommand*\mciteSetBstSublistLabelBeginEnd[3]{}
\providecommand*\EndOfBibitem{}
\mciteSetBstSublistMode{f}
\mciteSetBstMaxWidthForm{subitem}{(\alph{mcitesubitemcount})}
\mciteSetBstSublistLabelBeginEnd
  {\mcitemaxwidthsubitemform\space}
  {\relax}
  {\relax}

\bibitem[Mahadevan \latin{et~al.}(2004)Mahadevan, Zunger, and Sarma]{Mahadevan}
Mahadevan,~P.; Zunger,~A.; Sarma,~D.~D. Unusual directional dependence of
  exchange energies in GaAs diluted with Mn: Is the RKKY description relevant?
  \emph{Phys. Rev. Lett.} \textbf{2004}, \emph{93}, 177201\relax
\mciteBstWouldAddEndPuncttrue
\mciteSetBstMidEndSepPunct{\mcitedefaultmidpunct}
{\mcitedefaultendpunct}{\mcitedefaultseppunct}\relax
\EndOfBibitem
\bibitem[Sarma \latin{et~al.}(2000)Sarma, Mahadevan, Saha-Dasgupta, Ray, and
  Kumar]{Sarma}
Sarma,~D.~D.; Mahadevan,~P.; Saha-Dasgupta,~T.; Ray,~S.; Kumar,~A. Electronic
  structure of ${\mathrm{Sr}}_{2}{\mathrm{FeMoO}}_{6}$. \emph{Phys. Rev. Lett.}
  \textbf{2000}, \emph{85}, 2549--2552\relax
\mciteBstWouldAddEndPuncttrue
\mciteSetBstMidEndSepPunct{\mcitedefaultmidpunct}
{\mcitedefaultendpunct}{\mcitedefaultseppunct}\relax
\EndOfBibitem
\bibitem[Lee \latin{et~al.}(2016)Lee, Shin, Lee, and Lee]{Review1}
Lee,~J.~Y.; Shin,~J.~H.; Lee,~G.~H.; Lee,~C.~H. Two-dimensional semiconductor
  optoelectronics based on van der Waals heterostructures. \emph{Nanomaterials
  (Basel)} \textbf{2016}, \emph{6}, 193\relax
\mciteBstWouldAddEndPuncttrue
\mciteSetBstMidEndSepPunct{\mcitedefaultmidpunct}
{\mcitedefaultendpunct}{\mcitedefaultseppunct}\relax
\EndOfBibitem
\bibitem[Choi \latin{et~al.}(2017)Choi, Choudhary, Han, Park, Akinwande, and
  Lee]{Review2}
Choi,~W.; Choudhary,~N.; Han,~G.~H.; Park,~J.; Akinwande,~D.; Lee,~Y.~H. Recent
  development of two-dimensional transition metal dichalcogenides and their
  applications. \emph{Mater. Today} \textbf{2017}, \emph{20}, 116--130\relax
\mciteBstWouldAddEndPuncttrue
\mciteSetBstMidEndSepPunct{\mcitedefaultmidpunct}
{\mcitedefaultendpunct}{\mcitedefaultseppunct}\relax
\EndOfBibitem
\bibitem[Manzeli \latin{et~al.}(2017)Manzeli, Ovchinnikov, Pasquier, Yazyev,
  and Kis]{Review3}
Manzeli,~S.; Ovchinnikov,~D.; Pasquier,~D.; Yazyev,~O.; Kis,~A. 2D transition
  metal dichalcogenides. \emph{Nat. Rev. Mater.} \textbf{2017}, \emph{2},
  17033\relax
\mciteBstWouldAddEndPuncttrue
\mciteSetBstMidEndSepPunct{\mcitedefaultmidpunct}
{\mcitedefaultendpunct}{\mcitedefaultseppunct}\relax
\EndOfBibitem
\bibitem[Pal \latin{et~al.}(2017)Pal, Singh, G., Mahale, Kumar, Thirupathaiah,
  Sezen, Amati, Gregoratti, Waghmare, and Sarma]{BPal}
Pal,~B.; Singh,~A.; G.,~S.; Mahale,~P.; Kumar,~A.; Thirupathaiah,~S.;
  Sezen,~H.; Amati,~M.; Gregoratti,~L.; Waghmare,~U.~V. \latin{et~al.}
  Chemically exfoliated $\mathrm{Mo}{\mathrm{S}}_{2}$ layers: Spectroscopic
  evidence for the semiconducting nature of the dominant trigonal metastable
  phase. \emph{Phys. Rev. B} \textbf{2017}, \emph{96}, 195426\relax
\mciteBstWouldAddEndPuncttrue
\mciteSetBstMidEndSepPunct{\mcitedefaultmidpunct}
{\mcitedefaultendpunct}{\mcitedefaultseppunct}\relax
\EndOfBibitem
\bibitem[Parkin and Friend(1980)Parkin, and Friend]{Parkin_PhilMag1980}
Parkin,~S. S.~P.; Friend,~R.~H. 3d transition-metal intercalates of the niobium
  and tantalum dichalcogenides. I. Magnetic properties. \emph{Philos. Mag. B}
  \textbf{1980}, \emph{41}, 65--93\relax
\mciteBstWouldAddEndPuncttrue
\mciteSetBstMidEndSepPunct{\mcitedefaultmidpunct}
{\mcitedefaultendpunct}{\mcitedefaultseppunct}\relax
\EndOfBibitem
\bibitem[Parkin and Friend(1980)Parkin, and Friend]{Parkin_PhilMag1980_II}
Parkin,~S. S.~P.; Friend,~R.~H. 3d transition-metal intercalates of the niobium
  and tantalum dichalcogenides. II. Transport properties. \emph{Philos. Mag. B}
  \textbf{1980}, \emph{41}, 95--112\relax
\mciteBstWouldAddEndPuncttrue
\mciteSetBstMidEndSepPunct{\mcitedefaultmidpunct}
{\mcitedefaultendpunct}{\mcitedefaultseppunct}\relax
\EndOfBibitem
\bibitem[Inoue \latin{et~al.}(1985)Inoue, Matsumoto, Negishi, and
  Sakai]{InoueJMMM85}
Inoue,~M.; Matsumoto,~M.; Negishi,~H.; Sakai,~H. Low field ac magnetic
  susceptibility measurements of intercalation compounds M$_x$TiS$_2$ (M$=$3d
  transition metals). \emph{J. Magn. Magn. Mater.} \textbf{1985}, \emph{53},
  131--138\relax
\mciteBstWouldAddEndPuncttrue
\mciteSetBstMidEndSepPunct{\mcitedefaultmidpunct}
{\mcitedefaultendpunct}{\mcitedefaultseppunct}\relax
\EndOfBibitem
\bibitem[Negishi \latin{et~al.}(1987)Negishi, Shoube, Takahashi, Ueda, Sasaki,
  and Inoue]{NegishiJMMM1987}
Negishi,~H.; Shoube,~A.; Takahashi,~H.; Ueda,~Y.; Sasaki,~M.; Inoue,~M.
  Magnetic properties of intercalation compounds M$_x$TiS$_2$ (M = 3d
  transition metal). \emph{J. Magn. Magn. Mater.} \textbf{1987}, \emph{67},
  179--186\relax
\mciteBstWouldAddEndPuncttrue
\mciteSetBstMidEndSepPunct{\mcitedefaultmidpunct}
{\mcitedefaultendpunct}{\mcitedefaultseppunct}\relax
\EndOfBibitem
\bibitem[Inoue and Negishi(1986)Inoue, and Negishi]{Inoue_struct_JPhysChem86}
Inoue,~M.; Negishi,~H. Interlayer spacing of 3d transition-metal intercalates
  of 1T-CdI$_2$-type TiS$_2$. \emph{J. Phys. Chem.} \textbf{1986}, \emph{90},
  235--238\relax
\mciteBstWouldAddEndPuncttrue
\mciteSetBstMidEndSepPunct{\mcitedefaultmidpunct}
{\mcitedefaultendpunct}{\mcitedefaultseppunct}\relax
\EndOfBibitem
\bibitem[Koyano \latin{et~al.}(1986)Koyano, Negishi, Ueda, Sasaki, and
  Inoue]{Koyano_transp_physica86}
Koyano,~M.; Negishi,~H.; Ueda,~Y.; Sasaki,~M.; Inoue,~M. Electrical resistivity
  and thermoelectric power of intercalation compounds M$_x$TiS$_2$ (M$=$Mn, Fe,
  Co, and Ni). \emph{Phys. Status Solidi B} \textbf{1986}, \emph{138},
  357--363\relax
\mciteBstWouldAddEndPuncttrue
\mciteSetBstMidEndSepPunct{\mcitedefaultmidpunct}
{\mcitedefaultendpunct}{\mcitedefaultseppunct}\relax
\EndOfBibitem
\bibitem[Inoue \latin{et~al.}(1986)Inoue, Koyano, Negishi, and
  Ueda]{Inoue_transp_JPSJshort86}
Inoue,~M.; Koyano,~M.; Negishi,~H.; Ueda,~Y. Electronic properties of
  intercalation compound Fe$_x$TiS$_2$. \emph{J. Phys. Soc. Jpn.}
  \textbf{1986}, \emph{55}, 1400--1401\relax
\mciteBstWouldAddEndPuncttrue
\mciteSetBstMidEndSepPunct{\mcitedefaultmidpunct}
{\mcitedefaultendpunct}{\mcitedefaultseppunct}\relax
\EndOfBibitem
\bibitem[Negishi \latin{et~al.}(1988)Negishi, Koyano, Inoue, Sakakibara, and
  Goto]{NegishiJMMM88}
Negishi,~H.; Koyano,~M.; Inoue,~M.; Sakakibara,~T.; Goto,~T. High field
  magnetization of 3d transition metal intercalates M$_x$TiS$_2$ (M$=$3d
  metals). \emph{J. Magn. Magn. Mater.} \textbf{1988}, \emph{74}, 27--30\relax
\mciteBstWouldAddEndPuncttrue
\mciteSetBstMidEndSepPunct{\mcitedefaultmidpunct}
{\mcitedefaultendpunct}{\mcitedefaultseppunct}\relax
\EndOfBibitem
\bibitem[Yoshioka and Tazuke(1985)Yoshioka, and Tazuke]{YoshiokaJPSJ85}
Yoshioka,~T.; Tazuke,~Y. Magnetic properties of Fe$_x$TiS$_2$ system. \emph{J.
  Phys. Soc. Jpn.} \textbf{1985}, \emph{54}, 2088--2091\relax
\mciteBstWouldAddEndPuncttrue
\mciteSetBstMidEndSepPunct{\mcitedefaultmidpunct}
{\mcitedefaultendpunct}{\mcitedefaultseppunct}\relax
\EndOfBibitem
\bibitem[Negishi \latin{et~al.}(1988)Negishi, \={O}hara, Koyano, Inoue,
  Sakakibara, and Goto]{NegishiJPSJ88}
Negishi,~H.; \={O}hara,~S.; Koyano,~M.; Inoue,~M.; Sakakibara,~T.; Goto,~T.
  Anisotropic spin-glass and cluster-glass of layered Fe$_x$TiS$_2$ crystals.
  \emph{J. Phys. Soc. Jpn.} \textbf{1988}, \emph{57}, 4083--4085\relax
\mciteBstWouldAddEndPuncttrue
\mciteSetBstMidEndSepPunct{\mcitedefaultmidpunct}
{\mcitedefaultendpunct}{\mcitedefaultseppunct}\relax
\EndOfBibitem
\bibitem[Choe \latin{et~al.}(2019)Choe, Lee, Huang, Trivedi, and
  Morosan]{Choe_PRB19}
Choe,~J.; Lee,~K.; Huang,~C.-L.; Trivedi,~N.; Morosan,~E. Magnetotransport in
  Fe-intercalated $T{\mathrm{S}}_{2}$: Comparison between $T=\text{Ti}$ and Ta.
  \emph{Phys. Rev. B} \textbf{2019}, \emph{99}, 064420\relax
\mciteBstWouldAddEndPuncttrue
\mciteSetBstMidEndSepPunct{\mcitedefaultmidpunct}
{\mcitedefaultendpunct}{\mcitedefaultseppunct}\relax
\EndOfBibitem
\bibitem[Kuroiwa \latin{et~al.}(2000)Kuroiwa, Honda, and
  Noda]{Kuroiwa_neut_2000}
Kuroiwa,~Y.; Honda,~H.; Noda,~Y. Neutron magnetic scattering of intercalation
  compounds Fe$_x$TiS$_2$. \emph{Mol. Cryst. Liq. Cryst. Sci. Technol., Sect.
  A} \textbf{2000}, \emph{341}, 15--20\relax
\mciteBstWouldAddEndPuncttrue
\mciteSetBstMidEndSepPunct{\mcitedefaultmidpunct}
{\mcitedefaultendpunct}{\mcitedefaultseppunct}\relax
\EndOfBibitem
\bibitem[Chiew \latin{et~al.}(2020)Chiew, Miyata, Koyano, and
  Oshima]{FexTiS2_TEM_JPSJ2020}
Chiew,~Y.~L.; Miyata,~M.; Koyano,~M.; Oshima,~Y. Ordering of intercalated Fe
  atoms in Fe$_x$TiS$_2$ structures clarified using transmission electron
  microscopy. \emph{J. Phys. Soc. Jpn.} \textbf{2020}, \emph{89}, 074601\relax
\mciteBstWouldAddEndPuncttrue
\mciteSetBstMidEndSepPunct{\mcitedefaultmidpunct}
{\mcitedefaultendpunct}{\mcitedefaultseppunct}\relax
\EndOfBibitem
\bibitem[Suzuki \latin{et~al.}(1987)Suzuki, Yamasaki, and
  Motizuki]{Suzuki_bandcalc_JMMM87}
Suzuki,~N.; Yamasaki,~T.; Motizuki,~K. Electronic band structures of
  intercalation compounds of 3d transition metals with TiS$_2$. \emph{J. Magn.
  Magn. Mater.} \textbf{1987}, \emph{70}, 64--66\relax
\mciteBstWouldAddEndPuncttrue
\mciteSetBstMidEndSepPunct{\mcitedefaultmidpunct}
{\mcitedefaultendpunct}{\mcitedefaultseppunct}\relax
\EndOfBibitem
\bibitem[Suzuki \latin{et~al.}(1989)Suzuki, Yamasaki, and
  Motizuki]{Suzuki_bandcalc_JPSJ89}
Suzuki,~N.; Yamasaki,~T.; Motizuki,~K. Electronic band structures and bond
  orders of M$_{1/3}$TiS$_2$ (M$=$Mn, Fe, Co, Ni). \emph{J. Phys. Soc. Jpn.}
  \textbf{1989}, \emph{58}, 3280--3289\relax
\mciteBstWouldAddEndPuncttrue
\mciteSetBstMidEndSepPunct{\mcitedefaultmidpunct}
{\mcitedefaultendpunct}{\mcitedefaultseppunct}\relax
\EndOfBibitem
\bibitem[Martinez \latin{et~al.}(2002)Martinez, Tison, Baraille, Loudet, and
  Gonbeau]{Martinez_XPSSTMcalc_JESRP02}
Martinez,~H.; Tison,~Y.; Baraille,~I.; Loudet,~M.; Gonbeau,~D. Experimental
  (XPS/STM) and theoretical (FLAPW) studies of model systems M$_{1/4}$TiS$_2$
  (M$=$Fe, Co, Ni): influence of the inserted metal. \emph{J. Electron.
  Spectrosc. Relat. Phenom.} \textbf{2002}, \emph{125}, 181--196\relax
\mciteBstWouldAddEndPuncttrue
\mciteSetBstMidEndSepPunct{\mcitedefaultmidpunct}
{\mcitedefaultendpunct}{\mcitedefaultseppunct}\relax
\EndOfBibitem
\bibitem[Ueda \latin{et~al.}(1986)Ueda, Negishi, Koyano, Inoue, Soda, Sakamoto,
  and Suga]{Ueda_UPS_SSC86}
Ueda,~Y.; Negishi,~H.; Koyano,~M.; Inoue,~M.; Soda,~K.; Sakamoto,~H.; Suga,~S.
  Resonant photoemission studies of 3$d$ transition metal intercalates of
  TiS$_2$. \emph{Solid State Commun.} \textbf{1986}, \emph{57}, 839--842\relax
\mciteBstWouldAddEndPuncttrue
\mciteSetBstMidEndSepPunct{\mcitedefaultmidpunct}
{\mcitedefaultendpunct}{\mcitedefaultseppunct}\relax
\EndOfBibitem
\bibitem[Ueda \latin{et~al.}(1987)Ueda, Fukushima, Negishi, Inoue, Taniguchi,
  and Suga]{Ueda_UPS_JPSJ87}
Ueda,~Y.; Fukushima,~K.; Negishi,~H.; Inoue,~M.; Taniguchi,~M.; Suga,~S.
  Photoemission studies on intercalation compounds of M$_x$TiS$_2$ (M$=$3d
  transition metals). \emph{J. Phys. Soc. Jpn.} \textbf{1987}, \emph{56},
  2471--2476\relax
\mciteBstWouldAddEndPuncttrue
\mciteSetBstMidEndSepPunct{\mcitedefaultmidpunct}
{\mcitedefaultendpunct}{\mcitedefaultseppunct}\relax
\EndOfBibitem
\bibitem[Fujimori \latin{et~al.}(1988)Fujimori, Suga, Negishi, and
  Inoue]{Fujimori_XPS_PRB88}
Fujimori,~A.; Suga,~S.; Negishi,~H.; Inoue,~M. X-ray photoemission and
  Auger-electron spectroscopic study of the electronic structure of
  intercalation compounds ${M}_{x}$${\mathrm{TiS}}_{2}$ (M=Mn, Fe, Co, and Ni).
  \emph{Phys. Rev. B} \textbf{1988}, \emph{38}, 3676--3689\relax
\mciteBstWouldAddEndPuncttrue
\mciteSetBstMidEndSepPunct{\mcitedefaultmidpunct}
{\mcitedefaultendpunct}{\mcitedefaultseppunct}\relax
\EndOfBibitem
\bibitem[Suga(2000)]{Suga_UPSXAS_2000}
Suga,~S. Angle-resolved, resonance-and inverse-photoemission studies of
  transition metal intercalated TiS$_2$. \emph{Mol. Cryst. Liq. Cryst. Sci.
  Technol., Sect. A} \textbf{2000}, \emph{341}, 9--14\relax
\mciteBstWouldAddEndPuncttrue
\mciteSetBstMidEndSepPunct{\mcitedefaultmidpunct}
{\mcitedefaultendpunct}{\mcitedefaultseppunct}\relax
\EndOfBibitem
\bibitem[Yamasaki \latin{et~al.}(2002)Yamasaki, Imada, Sekiyama, Suga,
  Matsushita, Muro, Saitoh, Negishi, and Sasaki]{FexTiS2_YamasakiSuga}
Yamasaki,~A.; Imada,~S.; Sekiyama,~A.; Suga,~S.; Matsushita,~T.; Muro,~T.;
  Saitoh,~Y.; Negishi,~H.; Sasaki,~M. Angle-resolved photoemission spectroscopy
  and magnetic circular dichroism in Fe-intercalated TiS$_2$. \emph{Surf. Rev.
  Lett.} \textbf{2002}, \emph{9}, 961--966\relax
\mciteBstWouldAddEndPuncttrue
\mciteSetBstMidEndSepPunct{\mcitedefaultmidpunct}
{\mcitedefaultendpunct}{\mcitedefaultseppunct}\relax
\EndOfBibitem
\bibitem[Suga \latin{et~al.}(2015)Suga, Tusche, ichiro Matsushita, Ellguth,
  Irizawa, and Kirschner]{Suga_NixTiS2_ARPES_NJPhys2015}
Suga,~S.; Tusche,~C.; ichiro Matsushita,~Y.; Ellguth,~M.; Irizawa,~A.;
  Kirschner,~J. Momentum microscopy of the layered semiconductor TiS$_2$ and Ni
  intercalated Ni$_{1/3}$TiS$_2$. \emph{New J. Phys.} \textbf{2015}, \emph{17},
  083010\relax
\mciteBstWouldAddEndPuncttrue
\mciteSetBstMidEndSepPunct{\mcitedefaultmidpunct}
{\mcitedefaultendpunct}{\mcitedefaultseppunct}\relax
\EndOfBibitem
\bibitem[Tazuke \latin{et~al.}(2005)Tazuke, Ohta, and
  Miyamoto]{Tazuke_RKKY_JPSJ2005}
Tazuke,~Y.; Ohta,~Y.; Miyamoto,~S. Exchange interactions in Fe$_x$TiS$_2$.
  \emph{J. Phys. Soc. Jpn.} \textbf{2005}, \emph{74}, 2644--2645\relax
\mciteBstWouldAddEndPuncttrue
\mciteSetBstMidEndSepPunct{\mcitedefaultmidpunct}
{\mcitedefaultendpunct}{\mcitedefaultseppunct}\relax
\EndOfBibitem
\bibitem[Tazuke \latin{et~al.}(2006)Tazuke, Miyashita, Nakano, and
  Sasaki]{Tazuke_Physica06}
Tazuke,~Y.; Miyashita,~T.; Nakano,~H.; Sasaki,~R. Magnetic properties of
  M$_x$TiSe$_2$ (M$=$Mn, Fe, Co). \emph{Phys. Status Solidi C} \textbf{2006},
  \emph{3}, 2787--2790\relax
\mciteBstWouldAddEndPuncttrue
\mciteSetBstMidEndSepPunct{\mcitedefaultmidpunct}
{\mcitedefaultendpunct}{\mcitedefaultseppunct}\relax
\EndOfBibitem
\bibitem[L{\'{e}}vy(1979)]{SampleGrowth}
L{\'{e}}vy,~F., Ed. \emph{Intercalated Layered Materials}; D. Reidel Publishing
  Company: Dordrecht, Holland, 1979\relax
\mciteBstWouldAddEndPuncttrue
\mciteSetBstMidEndSepPunct{\mcitedefaultmidpunct}
{\mcitedefaultendpunct}{\mcitedefaultseppunct}\relax
\EndOfBibitem
\bibitem[Furuse \latin{et~al.}(2013)Furuse, Okano, Fuchino, Uchida, Fujihira,
  Fujihira, Kadono, Fujimori, and Koide]{vector_Furuse}
Furuse,~M.; Okano,~M.; Fuchino,~S.; Uchida,~A.; Fujihira,~J.; Fujihira,~S.;
  Kadono,~T.; Fujimori,~A.; Koide,~T. HTS vector magnet for magnetic circular
  dichroism measurement. \emph{IEEE Trans. Appl. Supercond.} \textbf{2013},
  \emph{23}, 4100704\relax
\mciteBstWouldAddEndPuncttrue
\mciteSetBstMidEndSepPunct{\mcitedefaultmidpunct}
{\mcitedefaultendpunct}{\mcitedefaultseppunct}\relax
\EndOfBibitem
\bibitem[Shibata \latin{et~al.}(2018)Shibata, Kitamura, Minohara, Yoshimatsu,
  Kadono, Ishigami, Harano, Takahashi, Sakamoto, Nonaka, Ikeda, Chi, Furuse,
  Fuchino, Okano, Fujihira, Uchida, Watanabe, Fujihira, Fujihira, Tanaka,
  Kumigashira, Koide, and Fujimori]{AngleDep_shibata}
Shibata,~G.; Kitamura,~M.; Minohara,~M.; Yoshimatsu,~K.; Kadono,~T.;
  Ishigami,~K.; Harano,~T.; Takahashi,~Y.; Sakamoto,~S.; Nonaka,~Y.
  \latin{et~al.}  Anisotropic spin-density distribution and magnetic anisotropy
  of strained La$_{1-x}$Sr$_x$MnO$_3$ thin films: angle-dependent x-ray
  magnetic circular dichroism. \emph{npj Quantum Mater.} \textbf{2018},
  \emph{3}, 3\relax
\mciteBstWouldAddEndPuncttrue
\mciteSetBstMidEndSepPunct{\mcitedefaultmidpunct}
{\mcitedefaultendpunct}{\mcitedefaultseppunct}\relax
\EndOfBibitem
\bibitem[Carra \latin{et~al.}(1993)Carra, Thole, Altarelli, and Wang]{SpinSum}
Carra,~P.; Thole,~B.~T.; Altarelli,~M.; Wang,~X. X-ray circular dichroism and
  local magnetic fields. \emph{Phys. Rev. Lett.} \textbf{1993}, \emph{70},
  694--697\relax
\mciteBstWouldAddEndPuncttrue
\mciteSetBstMidEndSepPunct{\mcitedefaultmidpunct}
{\mcitedefaultendpunct}{\mcitedefaultseppunct}\relax
\EndOfBibitem
\bibitem[Thole \latin{et~al.}(1992)Thole, Carra, Sette, and van~der
  Laan]{OrbSum}
Thole,~B.~T.; Carra,~P.; Sette,~F.; van~der Laan,~G. X-ray circular dichroism
  as a probe of orbital magnetization. \emph{Phys. Rev. Lett.} \textbf{1992},
  \emph{68}, 1943--1946\relax
\mciteBstWouldAddEndPuncttrue
\mciteSetBstMidEndSepPunct{\mcitedefaultmidpunct}
{\mcitedefaultendpunct}{\mcitedefaultseppunct}\relax
\EndOfBibitem
\bibitem[Teramura \latin{et~al.}(1996)Teramura, Tanaka, and Jo]{Teramura}
Teramura,~Y.; Tanaka,~A.; Jo,~T. Effect of Coulomb interaction on the x-ray
  magnetic circular dichroism spin sum rule in 3$d$ transition elements.
  \emph{J. Phys. Soc. Jpn.} \textbf{1996}, \emph{65}, 1053--1055\relax
\mciteBstWouldAddEndPuncttrue
\mciteSetBstMidEndSepPunct{\mcitedefaultmidpunct}
{\mcitedefaultendpunct}{\mcitedefaultseppunct}\relax
\EndOfBibitem
\bibitem[Piamonteze \latin{et~al.}(2009)Piamonteze, Miedema, and
  de~Groot]{Piamonteze}
Piamonteze,~C.; Miedema,~P.; de~Groot,~F. M.~F. Accuracy of the spin sum rule
  in XMCD for the transition-metal $L$ edges from manganese to copper.
  \emph{Phys. Rev. B} \textbf{2009}, \emph{80}, 184410\relax
\mciteBstWouldAddEndPuncttrue
\mciteSetBstMidEndSepPunct{\mcitedefaultmidpunct}
{\mcitedefaultendpunct}{\mcitedefaultseppunct}\relax
\EndOfBibitem
\bibitem[St\"ohr and K\"onig(1995)St\"ohr, and K\"onig]{TXMCD_Stohr}
St\"ohr,~J.; K\"onig,~H. Determination of spin- and orbital-moment anisotropies
  in transition metals by angle-dependent x-ray magnetic circular dichroism.
  \emph{Phys. Rev. Lett.} \textbf{1995}, \emph{75}, 3748--3751\relax
\mciteBstWouldAddEndPuncttrue
\mciteSetBstMidEndSepPunct{\mcitedefaultmidpunct}
{\mcitedefaultendpunct}{\mcitedefaultseppunct}\relax
\EndOfBibitem
\bibitem[D\"urr and van~der Laan(1996)D\"urr, and van~der Laan]{TXMCD_Durr}
D\"urr,~H.~A.; van~der Laan,~G. Magnetic circular x-ray dichroism in transverse
  geometry: Importance of noncollinear ground state moments. \emph{Phys. Rev.
  B} \textbf{1996}, \emph{54}, R760--R763\relax
\mciteBstWouldAddEndPuncttrue
\mciteSetBstMidEndSepPunct{\mcitedefaultmidpunct}
{\mcitedefaultendpunct}{\mcitedefaultseppunct}\relax
\EndOfBibitem
\bibitem[Tanaka and Jo(1994)Tanaka, and Jo]{TanakaCluster}
Tanaka,~A.; Jo,~T. Resonant 3$d$, 3$p$ and 3$s$ photoemission in transition
  metal oxides predicted at 2$p$ threshold. \emph{J. Phys. Soc. Jpn.}
  \textbf{1994}, \emph{63}, 2788--2807\relax
\mciteBstWouldAddEndPuncttrue
\mciteSetBstMidEndSepPunct{\mcitedefaultmidpunct}
{\mcitedefaultendpunct}{\mcitedefaultseppunct}\relax
\EndOfBibitem
\bibitem[Regan \latin{et~al.}(2001)Regan, Ohldag, Stamm, Nolting, L\"uning,
  St\"ohr, and White]{FeXASref_Regan}
Regan,~T.~J.; Ohldag,~H.; Stamm,~C.; Nolting,~F.; L\"uning,~J.; St\"ohr,~J.;
  White,~R.~L. Chemical effects at metal/oxide interfaces studied by
  x-ray-absorption spectroscopy. \emph{Phys. Rev. B} \textbf{2001}, \emph{64},
  214422\relax
\mciteBstWouldAddEndPuncttrue
\mciteSetBstMidEndSepPunct{\mcitedefaultmidpunct}
{\mcitedefaultendpunct}{\mcitedefaultseppunct}\relax
\EndOfBibitem
\bibitem[Chen \latin{et~al.}(1995)Chen, Idzerda, Lin, Smith, Meigs, Chaban, Ho,
  Pellegrin, and Sette]{CTChen}
Chen,~C.~T.; Idzerda,~Y.~U.; Lin,~H.-J.; Smith,~N.~V.; Meigs,~G.; Chaban,~E.;
  Ho,~G.~H.; Pellegrin,~E.; Sette,~F. Experimental confirmation of the x-ray
  magnetic circular dichroism sum rules for iron and cobalt. \emph{Phys. Rev.
  Lett.} \textbf{1995}, \emph{75}, 152--155\relax
\mciteBstWouldAddEndPuncttrue
\mciteSetBstMidEndSepPunct{\mcitedefaultmidpunct}
{\mcitedefaultendpunct}{\mcitedefaultseppunct}\relax
\EndOfBibitem
\bibitem[Kimura \latin{et~al.}(1993)Kimura, Suga, Matsushita, Imada, Shino,
  Saitoh, Shigeoka, Daimon, Kinoshita, Kakizaki, Oh, Negishi, and
  Inoue]{Kimura_TiS2}
Kimura,~A.; Suga,~S.; Matsushita,~T.; Imada,~S.; Shino,~N.; Saitoh,~Y.;
  Shigeoka,~H.; Daimon,~H.; Kinoshita,~T.; Kakizaki,~A. \latin{et~al.}
  Electronic structure of Ni intercalated TiS$_2$ probed by angle resolved and
  2$p$ core resonance photoemission as well as by 2$p$ core absorption
  spectroscopy. \emph{Jpn. J. Appl. Phys.} \textbf{1993}, \emph{32},
  255--257\relax
\mciteBstWouldAddEndPuncttrue
\mciteSetBstMidEndSepPunct{\mcitedefaultmidpunct}
{\mcitedefaultendpunct}{\mcitedefaultseppunct}\relax
\EndOfBibitem
\bibitem[Cao \latin{et~al.}(2016)Cao, Liu, Kareev, Choudhury, Middey, Meyers,
  Kim, Ryan, Freeland, and Chakhalian]{TiXASref_Cao}
Cao,~Y.; Liu,~X.; Kareev,~M.; Choudhury,~D.; Middey,~S.; Meyers,~D.;
  Kim,~J.-W.; Ryan,~P.; Freeland,~J.; Chakhalian,~J. Engineered Mott ground
  state in a LaTiO$_{3+\delta}$/LaNiO$_3$ heterostructure. \emph{Nat. Commun.}
  \textbf{2016}, \emph{7}, 10418\relax
\mciteBstWouldAddEndPuncttrue
\mciteSetBstMidEndSepPunct{\mcitedefaultmidpunct}
{\mcitedefaultendpunct}{\mcitedefaultseppunct}\relax
\EndOfBibitem
\bibitem[Stoner and Wohlfarth(1948)Stoner, and Wohlfarth]{SWmodel}
Stoner,~E.~C.; Wohlfarth,~E.~P. A mechanism of magnetic hysteresis in
  heterogeneous alloys. \emph{Philos. Trans. R. Soc., A} \textbf{1948},
  \emph{240}, 599--642\relax
\mciteBstWouldAddEndPuncttrue
\mciteSetBstMidEndSepPunct{\mcitedefaultmidpunct}
{\mcitedefaultendpunct}{\mcitedefaultseppunct}\relax
\EndOfBibitem
\bibitem[Bocquet \latin{et~al.}(1992)Bocquet, Mizokawa, Saitoh, Namatame, and
  Fujimori]{Bouquet_PRB1991}
Bocquet,~A.~E.; Mizokawa,~T.; Saitoh,~T.; Namatame,~H.; Fujimori,~A. Electronic
  structure of 3$d$-transition-metal compounds by analysis of the 2$p$
  core-level photoemission spectra. \emph{Phys. Rev. B} \textbf{1992},
  \emph{46}, 3771--3784\relax
\mciteBstWouldAddEndPuncttrue
\mciteSetBstMidEndSepPunct{\mcitedefaultmidpunct}
{\mcitedefaultendpunct}{\mcitedefaultseppunct}\relax
\EndOfBibitem
\end{mcitethebibliography}

\end{document}